\documentclass[10pt, conference]{FMFP2024}
\usepackage[top=1.5cm, bottom=1.5cm, left=1.884cm, right=1.884cm, includehead, includefoot, heightrounded]{geometry}

\usepackage{graphicx}
\usepackage{epstopdf, epsfig}
\usepackage{amsmath}

\usepackage{subfigure}
\usepackage{color}
\usepackage{float} 
\usepackage{url}
\usepackage{fancyhdr}
\usepackage{textcomp}

\renewcommand{\headrulewidth}{0pt} 

\begin{document}

\title{\LARGE \bf Influence of Aspect ratio in the Convection in Rotating Annulus In the Presence of Localized Heating}

\author{\textbf{Ayan Kumar Banerjee\textsuperscript{1,*}, Shivam Swarnakar\textsuperscript{2}}\\\\
\textsuperscript{1}\small{School of Artificial Intelligence, Amrita Vishwa Vidyapeetham,  Coimbatore, Tamil Nadu, India}\\
\textsuperscript{2} \small{Department of Mechanical Engineering, Indian Institute of Technology, India}\\
\small{*Corresponding Author, email: ayanbanerjee1@gmail.com}}

\vspace{0.3cm}

\maketitle
\thispagestyle{fancy} 

\begin{abstract}
Two-dimensional (2D) axisymmetric simulations are performed to study convection in a rotating cylindrical annulus subjected to localized heating. The flow is driven by a peripheral heating source located near the outer edge of the bottom surface, while the inner cylindrical boundary is uniformly cooled, generating a radially directed temperature gradient that drives buoyancy-induced motion. The localized heating arrangement establishes both radial and vertical thermal gradients near the outer boundary, reproducing thermal stratification characteristics that are representative of many atmospheric circulation systems. The convective dynamics are examined across a range of aspect ratios \((\Gamma)\), spanning Rayleigh numbers \(Ra = 2.4 \times 10^{7}\) to \(1.2 \times 10^{9}\) and Taylor numbers \(Ta = 1.6 \times 10^{7}\) to \(1.2 \times 10^{9}\), including the non-rotating case \((Ta = 0)\). Convection is predominantly restricted to narrow boundary layers, while the fluid interior remains diffusion dominated. In the non-rotating case, the temperature field is characterized by nearly horizontal isotherms. Under rotational forcing, however, the establishment of quasi-hydrostatic and geostrophic balances within the bulk fluid facilitates the redistribution of heat, resulting in a pronounced spreading of the isotherms across the interior region. The Nusselt number (\(Nu\)) depends sensitively on the governing parameters \(Ra\), \(Ta\), and \(\Gamma\). At moderate and high \(Ra\), heat transport obeys the scaling \(Nu \sim Ra^{1/4}\) and is largely unaffected by rotation. However, at low \(Ra\) and high \(Ta\), rotational effects suppress buoyancy-driven motion, leading to a substantial decrease in \(Nu\). The aspect ratio \(\Gamma\) also exerts a strong influence on heat transfer: \(Nu\) increases rapidly at low \(\Gamma\) due to confinement-enhanced convection, whereas for \(\Gamma > 1\), the growth rate decreases as the convective structures approach saturation and the flow becomes less constrained by geometry. Furthermore, the ratio of thermal to Ekman boundary layer thickness \((\delta_T/\delta_E)\) governs heat transfer efficiency: when \(\delta_E < \delta_T\), high rotation diminishes \(Nu\), whereas when \(\delta_E > \delta_T\), bulk mixing mitigates rotational suppression, resulting in a relatively stable \(Nu\).

\end{abstract}


\section{\bf{INTRODUCTION}}

\begin{figure*}
	\centering
                     \subfigure[]{%
		\includegraphics[scale=0.65]{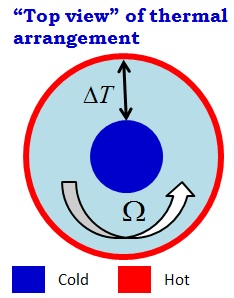}
		\label{$t$=82$s$}}
	\qquad   
	\subfigure[]{%
		\includegraphics[scale=0.45]{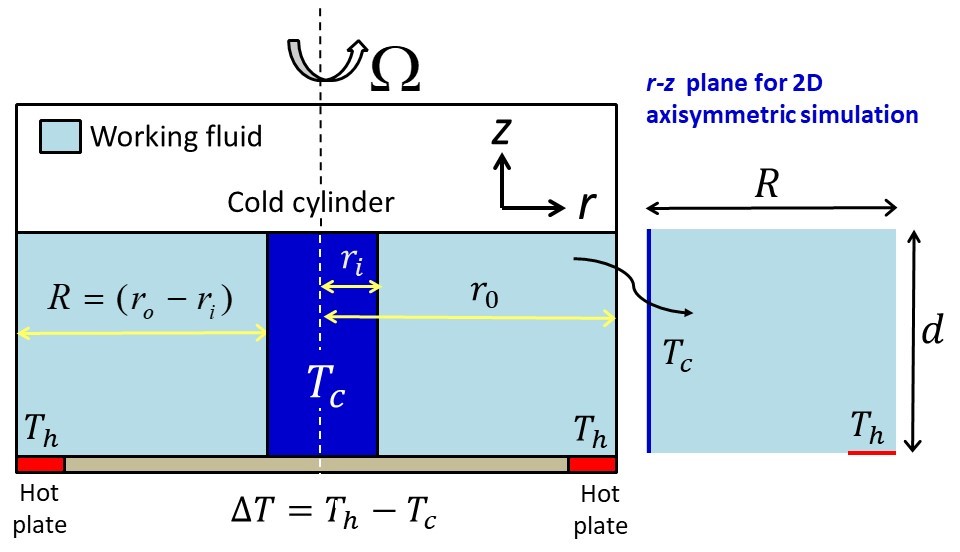}
		\label{$t$=102$s$}}
	\caption{\label{fluent_schematic} Schematic illustrating the experimental setup. (a) Top view perspective displaying the heating arrangement on $z$ plane (Banerjee et al.~\cite{Banerjee2024}). (b) Schematic depicting the $r-z$ plane utilized for the 2D axisymmetric simulation.}
	\label{fig:figure}
\end{figure*}

The interaction between buoyancy forces generated by spatially non-uniform thermal forcing and Coriolis forces gives rise to the intriguing fluid dynamical phenomenon known as rotating convection. This phenomenon is especially significant in the  context of Geophysical fluid dynamics, particularly in understanding Atmospheric circulation. Consequently, the rotating annulus, in which fluid is heated near the outer wall and cooled at the inner wall, serves as a canonical laboratory model for investigating the fundamental dynamics of geophysical and planetary atmospheres. Although this setup can replicate flow structures characteristic of the mid-latitude baroclinic zone in Earth’s atmosphere, However, attaining a comprehensive quantitative understanding of the mechanisms responsible for the formation of statically stable but baroclinically unstable regions in Earth's mid-latitudes remains a major scientific challenge. This limitation arises because the classical baroclinic annulus, with isothermal walls and uni-directional forcing, restricts baroclinic–stratification studies and poorly represents atmospheric systems where bi-directional forcing—both vertical and radial (meridional) thermal gradients dominate~\cite{Banerjee2018}. Banerjee et al.~\cite{Banerjee2016,Banerjee2018,Banerjee2021} introduced a novel bi-directionally forced rotating convection experiment, a modified form of the classical baroclinic annulus, consisting of a rotating fluid annulus with uniform cooling at the inner wall and localized heating at the outer base through a thin aluminum strip [Fig. \ref{fluent_schematic}(a)]. This configuration provides a unique platform for investigating the interaction between baroclinic wave dynamics and the background stratification (\cite{Banerjee2016,Banerjee2018,Banerjee2021, banerjee2016iccms,banerjee2018thermacomp,Swarnakar2021,Banerjee2025a,Banerjee2025b,nandan2026floquet, suresh2026rom}).


Banerjee et al.~\cite{Banerjee2018} investigated this configuration and demonstrated the coexistence and interaction of columnar convective plumes (CCPs) near the outer boundary with baroclinic waves—either steady or irregular depending on the Taylor number, $Ta$—within the fluid interior. Their study revealed that these flow structures collectively govern the thermal and velocity fields. In a subsequent investigation, Banerjee et al.~\cite{Banerjee2021} quantified the dependence of the Nusselt number ($Nu$) on the Taylor number ($Ta$) and the imposed heating rate ($Q$). Their results showed that heat transfer is strongly influenced by buoyancy forcing, whereas its dependence on the rotation rate is comparatively weak. Building on earlier work, Banerjee~\cite{Banerjee2024} performed two-dimensional (2D) axisymmetric simulations of the proposed system to examine its local heat transfer dynamics. These simulations employed a 12 cm × 12 cm computational domain, corresponding to an aspect ratio of \( \Gamma= d / R= 1 \), using geometric dimensions and fluid properties consistent with prior laboratory experiments (~\cite{Banerjee2018,Banerjee2021}). Here, \( R = (r_o - r_i) \) represents the annulus gap, and \( d \) is the fluid layer depth. The $\Gamma =1$ configuration, where  horizontal and vertical scales are comparable, served as a reference. The present study extends these axisymmetric simulations to investigate how variations in the aspect ratio, $\Gamma= d / R$ influence  convection pattern, both local and global dynamics of the system.

\begin{table*}[]
\caption{Parameter Space Considered in the Numerical Simulations}
\label{Tab:res3_1}
\centering
\small
\renewcommand{\arraystretch}{1.1}
\setlength{\tabcolsep}{8pt}
\begin{tabular}{lllllll}
\hline\noalign{\smallskip}
\qquad $\textbf{Parameter}$ & \quad  $\textbf{Symbols}$ & \quad \qquad $\textbf{Values}$ &  $\textbf{Units}$ \\
\noalign{\smallskip}\hline\noalign{\smallskip}
$$ Radial temperature gap $$ &\qquad  $\Delta$ T &\qquad \quad 0.5-50 &   K \\
$$ Background rotation rate &\qquad  $\Omega$ &\qquad \quad 0 -1.36 &  rad s$^{-1}$\\
$$ Density $$ &\qquad  $\rho$ &\qquad  \quad 998.2 &  kg m$^{-3}$\\
$$ Kinematic viscosity $$ &\qquad  $\nu$ &\qquad \quad $1.004\times10^{-6}$ &  m$^{2}$ s$^{-1}$\\
$$ Thermal diffusivity $$ &\qquad  $\kappa$ &\qquad \quad $0.143\times10^{-6}$ & m$^{2}$ s$^{-1}$\\
$$ Coeff. of volumetric thermal\\ expansion $$ &\qquad $\beta$ &\qquad \quad 0.000207 &  K$^{-1}$\\
$$ Inner radius $$ &\quad \quad  $\textit r_{i}$ &\qquad \quad 0.075 & m \\
$$ Outer radius $$ &\qquad  $\textit r_{o}$ &\qquad \quad 0.195 & m\\
$$ Annulus gap&\quad  $R=(r_{o}-r_{i})$ &\qquad \quad 0.12 & m \\
$$ Fluid height &\quad \quad  $\textit d$ &\quad \quad 0.012 - 0.36 &  m \\
$$ Aspect ratio $$ &\quad $\Gamma=\frac{d}{R}$ &\quad \qquad 0.1 - 3 \\
$$ Rayleigh number $$ &\qquad  $Ra$ &\quad $1.2\times10^{7}$ - $1.2\times10^{9}$ \\
$$ Taylor number $$ &\qquad  $Ta$ &\qquad 0 - $1.5\times10^{9}$ \\
$$ Ekman number $$ &\qquad  $Ek$ & \quad \quad $5.1\times 10^{-5}$ - $\infty$ \\
$$ Prandtl number $$ &\qquad  $Pr$ &\qquad \qquad 7&\qquad \\
\noalign{\smallskip}\hline
\end{tabular}
\end{table*}

Simulations using ANSYS-Fluent were conducted to study equilibrated 2D axisymmetric flows across the parameter spaces, aspect ratio $\Gamma \in [0.1  \quad 3]$,  Taylor number $Ta=\frac{4\Omega^2 R^4}{\nu^2}  \in [0 - 1.5\times10^{9}]$, Rayleigh number $Ra=\frac{\beta g \Delta T R^3}{\nu \kappa}  \in [1.2\times10^{7} - 1.2\times10^{9}]$, and Prandtl number $Pr=\frac{\nu}{\kappa}=7$.  Here,  $\nu$ represents kinematic viscosity, $\kappa$ denotes thermal diffusivity, $\Omega$ denotes the background rotation, $\beta$ stands for the coefficient of thermal expansion, $\Delta T$ signifies the radial temperature differential. The domain (Fig. \ref{fluent_schematic}(b)) represents the $r-z$ plane of the experimental setup in Banerjee et al.~\cite{Banerjee2018,Banerjee2021} with vertical boundaries as inner/outer radii and horizontal boundaries as the top/bottom lids of the cylindrical annulus (Fig. \ref{fluent_schematic}). 

\begin{figure*}
\centering
\subfigure[]{%
\includegraphics[scale=0.38]{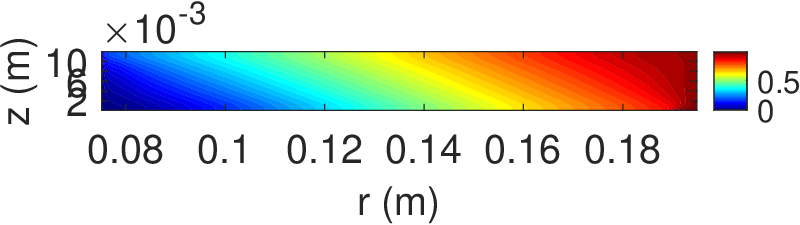}}
\subfigure[]{%
\includegraphics[scale=0.48]{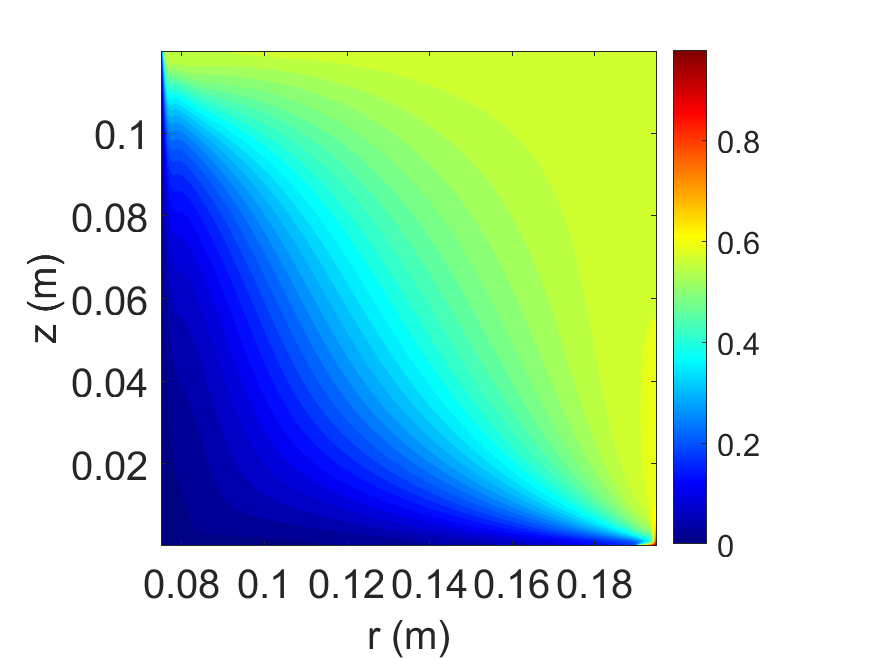}}
\subfigure[]{%
\includegraphics[scale=0.75]{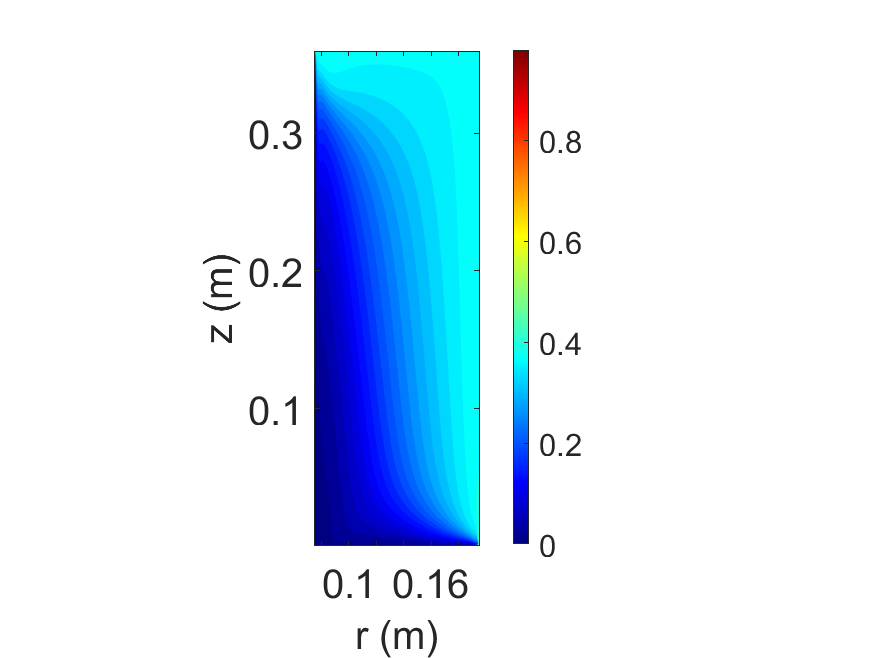}}
\caption{\label{fig:AR_norm_Temperature_contours} Normalised temperature $(T^{*})$ contours for (a) $\Gamma = 0.1$, (b) $\Gamma = 1$, (c) $\Gamma = 3$. The Rayleigh number $(Ra)$ and Taylor number $(Ta)$ are $5.9 \times 10^{8}$  and $1.52 \times 10^{9}$ respectively. Normalised temperature is given as: $T^{*} = (T - T_{c})/\Delta T $, where $T_{c}$ is the temperature of the cold wall and $\Delta T$ is the applied temperature gradient. Figures are not to the scale.} 
\end{figure*}

\begin{figure}[]
\centering
\includegraphics[width=0.8 \linewidth]{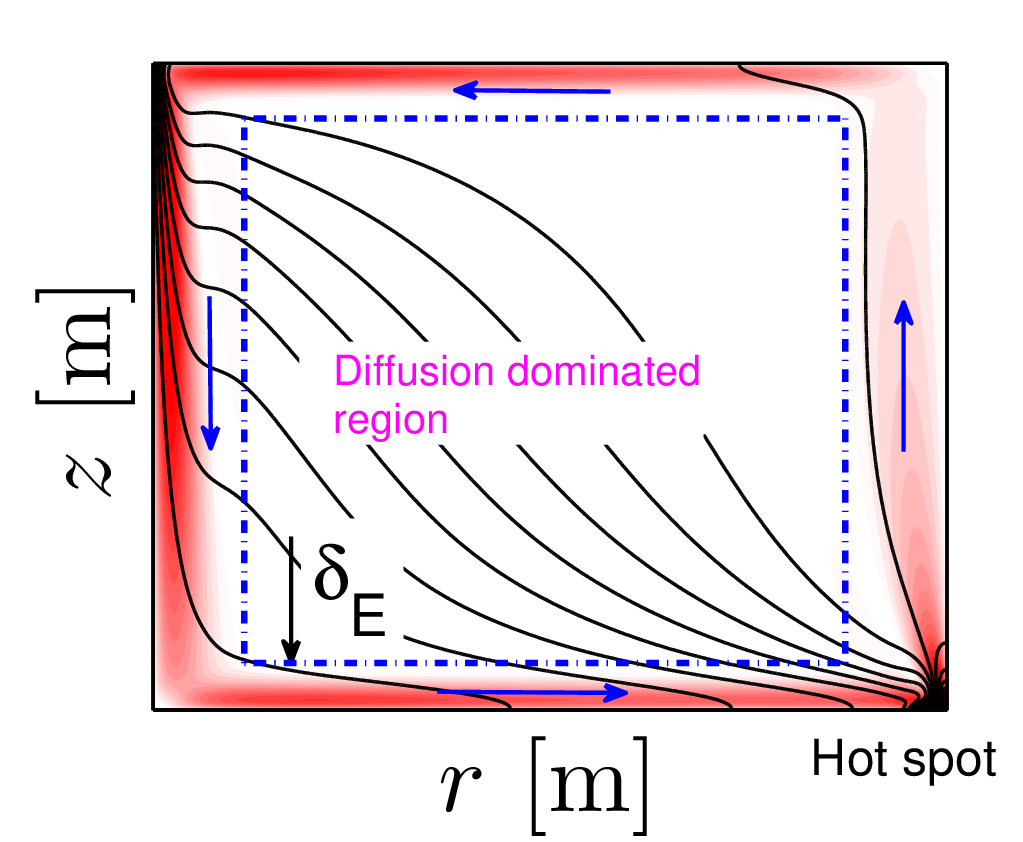}
\caption{\label{flow_schematic_simu} Schematic illustrating the flow dynamics in the rotating annulus. Filled contours represent the normalized vertical-plane velocity, $u_{vert}^n=(u_{\text{vert}} - u_{\text{vert}}^{\text{min}})/(u_{\text{vert}}^{\text{max}} - u_{\text{vert}}^{\text{min}})$
where \(u_{\mathrm{vert}}^{\min}\) and \(u_{\mathrm{vert}}^{\max}\) denote the minimum and maximum values of \(u_{\mathrm{vert}}\), respectively. The dimensional vertical-plane velocity is defined as $u_{vert}=\sqrt{u_r^2+u_z^2}$
with \(u_r\) and \(u_z\) representing the radial and axial velocity components in the rotating reference frame. Solid black lines indicate contours of the normalized temperature field, \(T^{*}\), while blue arrows highlight the direction of fluid motion within the boundary layers. The figure demonstrates that the fluid core remains largely quiescent in the \(r\)-\(z\) plane, whereas the strongest circulation is confined to the near-wall boundary-layer regions.}
\label{fig:sample1}
\end{figure}

\begin{figure}
\centering
\subfigure[]{%
\includegraphics[scale=0.33]{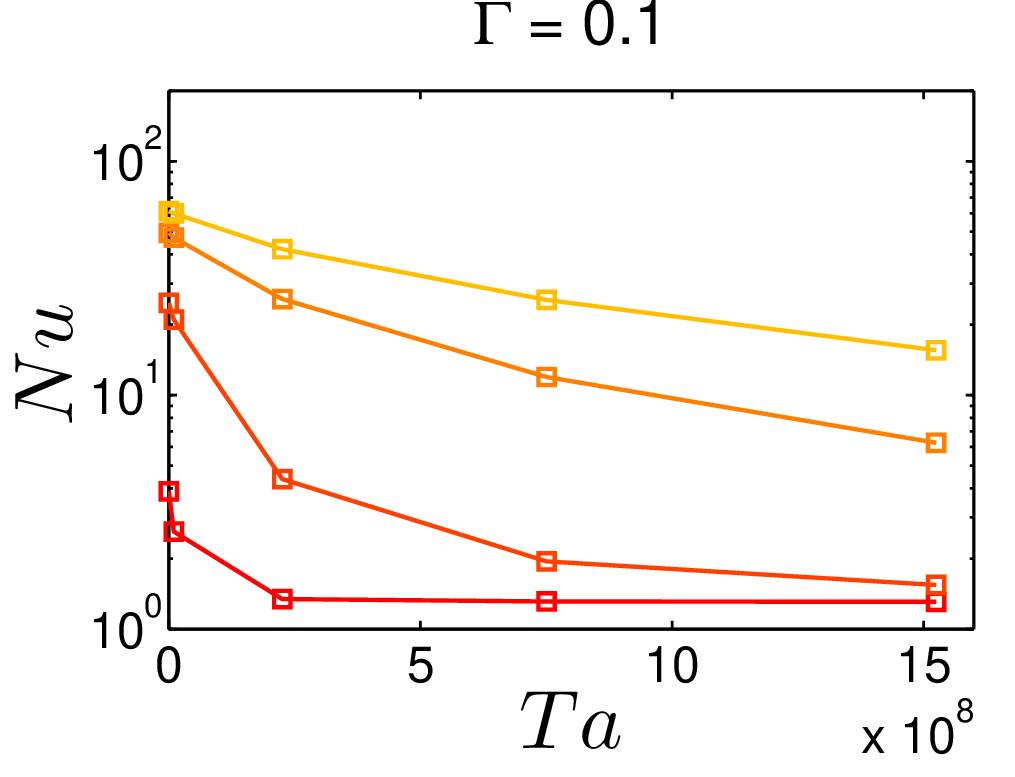}
\label{$t$=82$s$}}
\subfigure[]{%
\includegraphics[scale=0.33]{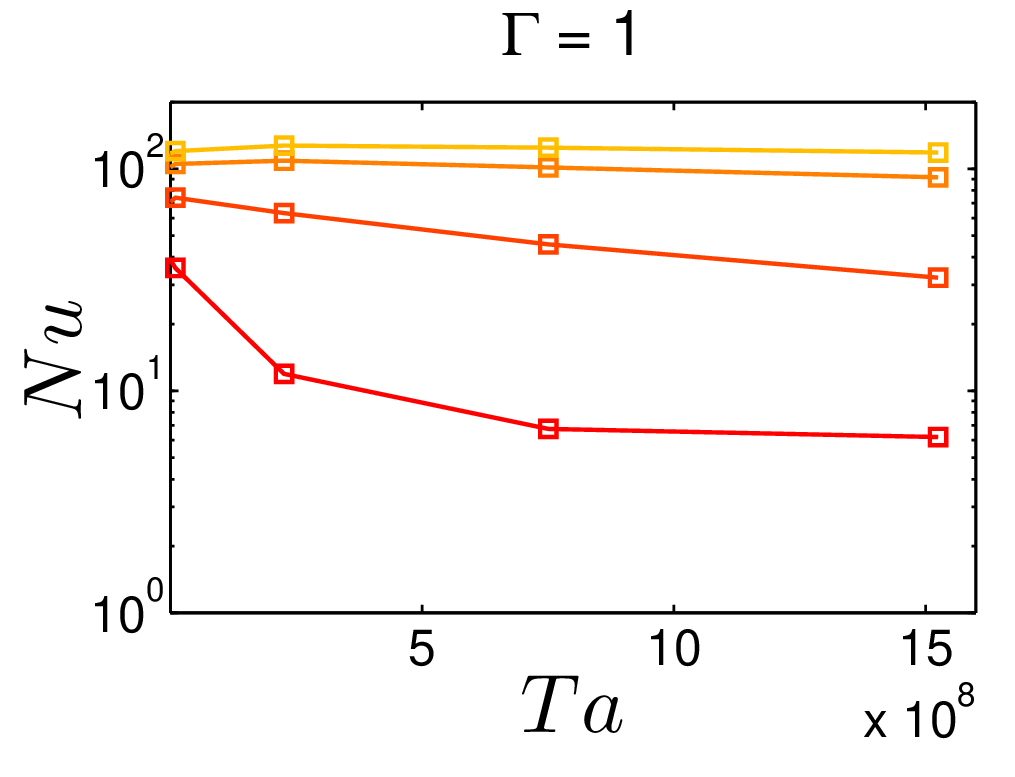}
\label{$t$=82$s$}}
\subfigure[]{%
\includegraphics[scale=0.37]{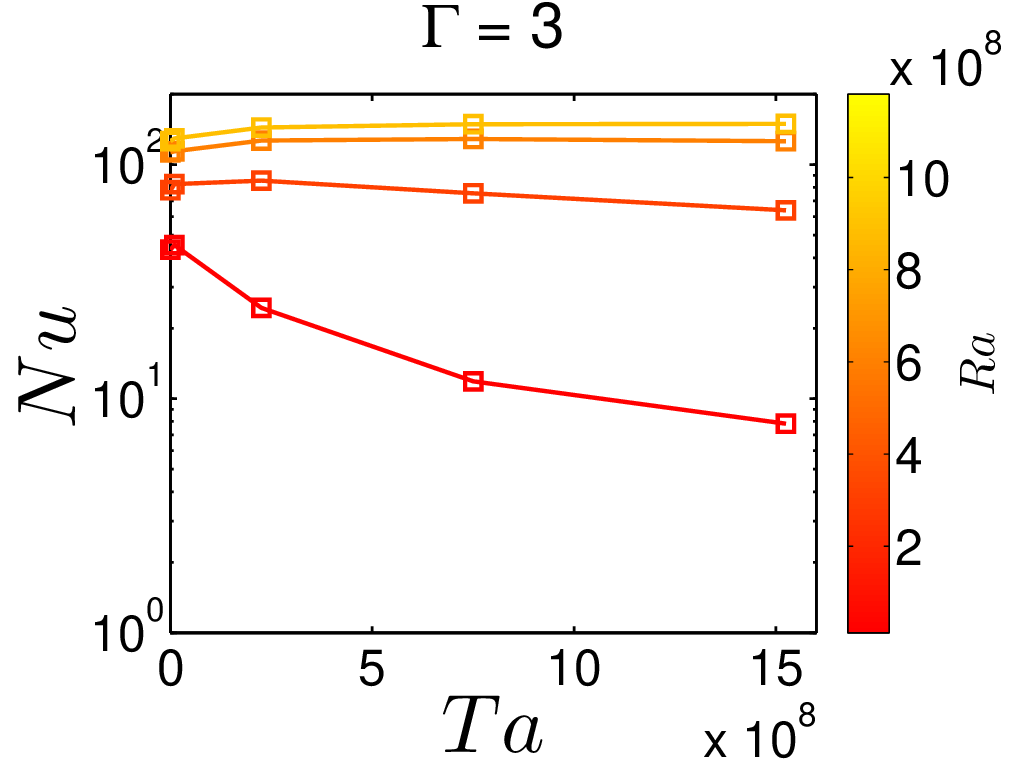}
\label{$t$=102$s$}}
\caption{\label{Nu_vs_Ta_AR} Nusselt number variation with $Ta$ at different aspect ratio. (a) $\Gamma=0.1$, (b) $\Gamma=1$, and (c) $\Gamma=3$.}
\label{fig:figure}
\end{figure} 




\section{Numerical Methodology}\label{sec-2}



Assuming axisymmetry \(\left(\frac{\partial}{\partial \theta}=0\right)\), incompressible flow, and the Boussinesq approximation (where density variations are neglected everywhere except in the buoyancy term), the continuity, Navier--Stokes, and energy equations in cylindrical coordinates \((r,\theta,z)\) can be expressed using Einstein's summation convention as follows:

\begin{equation}\label{eq:res3_1}
\frac{1}{r}\frac{\partial (ru_r)}{\partial r}+\frac{\partial u_z}{\partial z} = 0
\end{equation}
\begin{multline}\label{eq:momentum}
\frac{\partial u_i}{\partial t} + u_j \frac{\partial u_i}{\partial x_j} + \delta_{i\theta} \left( \frac{u_r u_\theta}{r} - 2\Omega u_r \right) 
- \delta_{ir} \frac{u_\theta^2}{r} + \delta_{iz} \beta g (T - T_0) 
\\= -\frac{\partial p}{\partial x_i} 
+ \nu \left[ \frac{\partial^2 u_i}{\partial x_j^2} + \frac{1}{r} \frac{\partial}{\partial r} \left( r \frac{\partial u_i}{\partial r} \right) - \delta_{i\theta} \frac{u_i}{r^2} \right]
\end{multline}
\begin{align}\label{eq:res3_5}
\frac{\partial T}{\partial t} + u_j \frac{\partial T}{\partial x_j} &=
\kappa \left[ \frac{\partial^2 T}{\partial x_j^2} + \frac{1}{r} \frac{\partial}{\partial r} \left( r \frac{\partial T}{\partial r} \right) \right]
\end{align}

where: $i, j = r, \theta, z$, $u_i = (u_r, u_\theta, u_z)$ are the relative radial,  azimuthal and axial (or vertical) velocity fields respectively, $x_i = (r, \theta, z)$ are the spatial coordinates, $\delta_{ij}$ is the Kronecker delta, $T_0$ is a reference temperature, $p$ signifies the kinematic pressure field, and $T$ denotes the temperature field. Centrifugal acceleration is neglected.

A constant cold temperature, \(T=T_c\), is maintained at the inner cylindrical boundary, while a constant hot temperature, \(T=T_h\), is prescribed over the localized heating plate. Numerical simulations are performed using a pressure-based solver with coupled pressure--velocity coupling. Since the flow remains within the low-Reynolds-number regime, a laminar flow model is adopted. The convective and diffusive terms are discretized using second-order upwind and second-order central-difference schemes, respectively. Fluid properties are assumed invariant, and suitable under-relaxation factors are employed to ensure robust convergence and numerical stability throughout the computations.

\section{\bf{RESULTS}}\label{Results}


\vspace{0.3cm}

\subsection{\textbf{Structure of thermal and velocity fields:}}

The flow and thermal characteristics are first investigated through the normalized temperature field in the \(r\)-\(z\) plane. The corresponding dimensionless temperature, \(T^{*}(r,z)\), is defined as

\begin{equation}\label{eq:T_norm}
T^{*}(r, z)=\frac{T(r, z)-T_c}{\Delta T}
\end{equation}

Here, \(T(r,z)\) denotes the dimensional temperature field, \(T_c\) is the temperature of the cold inner wall, and \(\Delta T = T_h - T_c\) is the imposed temperature difference. The normalized temperature contours reveal the development of a buoyancy-driven convective plume above the localized heating zone, consistent with the experimental findings of Banerjee et al.~\cite{Banerjee2018,Banerjee2021}. Strong temperature gradients are confined primarily to the vicinity of the heated plate, suggesting that heat transfer is dominated by localized thermal activity in this region. A distinct thermal boundary layer is also observed along the cold inner wall, characterized by a sharp radial temperature gradient separating the wall from the relatively well-mixed fluid core.

The normalised temperature contours $(T^*)$ for various values of aspect ratio $(\Gamma)$ are shown in Figure. \ref{fig:AR_norm_Temperature_contours} at $Ra = 5.9 \times 10^{8}$ and $Ta = 1.52 \times 10^{9}$. The isotherms are nearly vertical for $\Gamma = 3$ and maximally tilted for $\Gamma = 0.5$ [Figure.  \ref{fig:AR_norm_Temperature_contours}(a)-(c)]. This trend arises from the annulus geometry for fixed \(Ra\) and \(Ta\). Higher \(\Gamma\) values correspond to lower fluid temperatures, likely because a larger cooling wall length enhances heat loss. For example, at \(\Gamma = 3\) (largest cooling wall length), the fluid temperature is lowest. Similar temperature structures appear across all tested \(Ra\) and \(Ta\) values. The variation of isotherm slope,  
\[
S = \frac{\int_0^d \int_{r_{i}}^{r_{0}} \left(\frac{\partial T}{\partial r}\bigg/\frac{\partial T}{\partial z}\right) r\, dr\, dz}{\int_0^d \int_{r_{i}}^{r_{0}} r\, dr\, dz},
\]  
with \(\Gamma\) is identified as a promising avenue for future study.

\begin{figure*}
\centering
\begin{tabular}{c c c}
\subfigure[]{%
\includegraphics[scale=0.3]{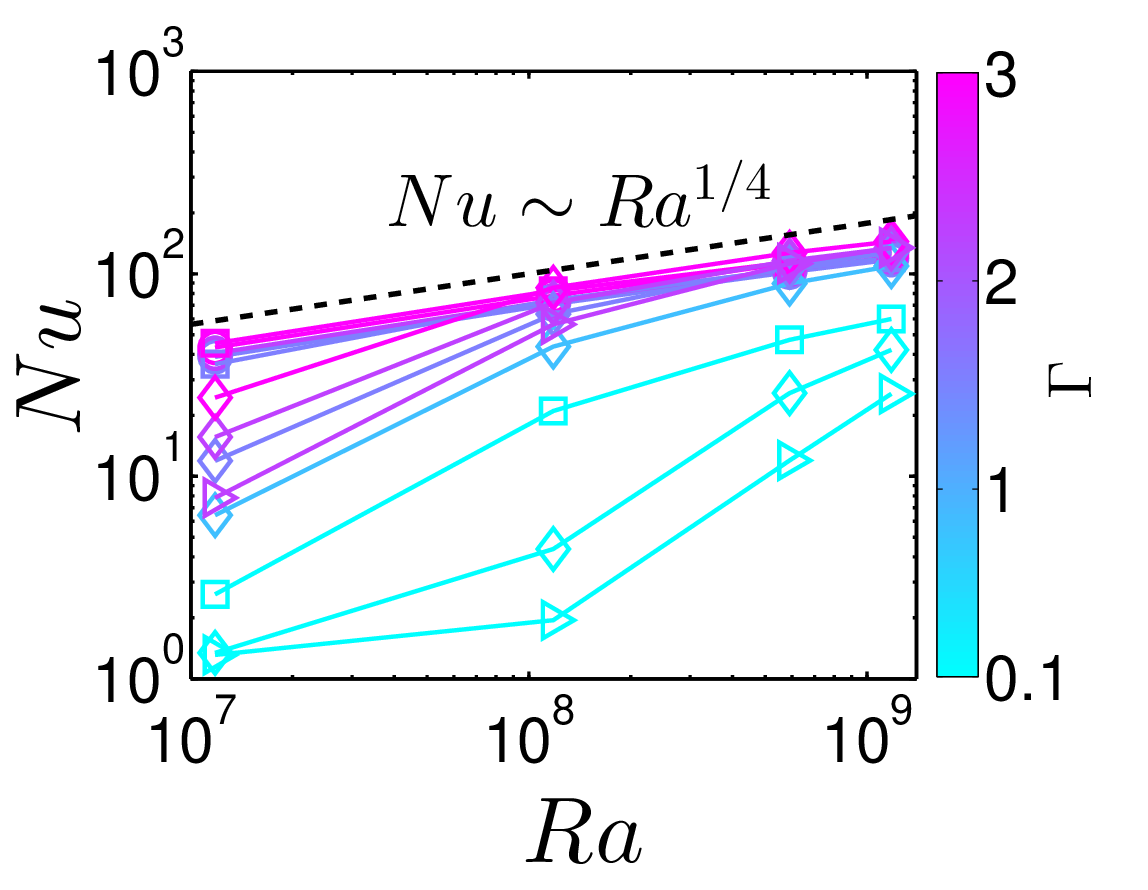}
\label{$t$=102$s$}}
\subfigure[]{%
\includegraphics[scale=0.3]{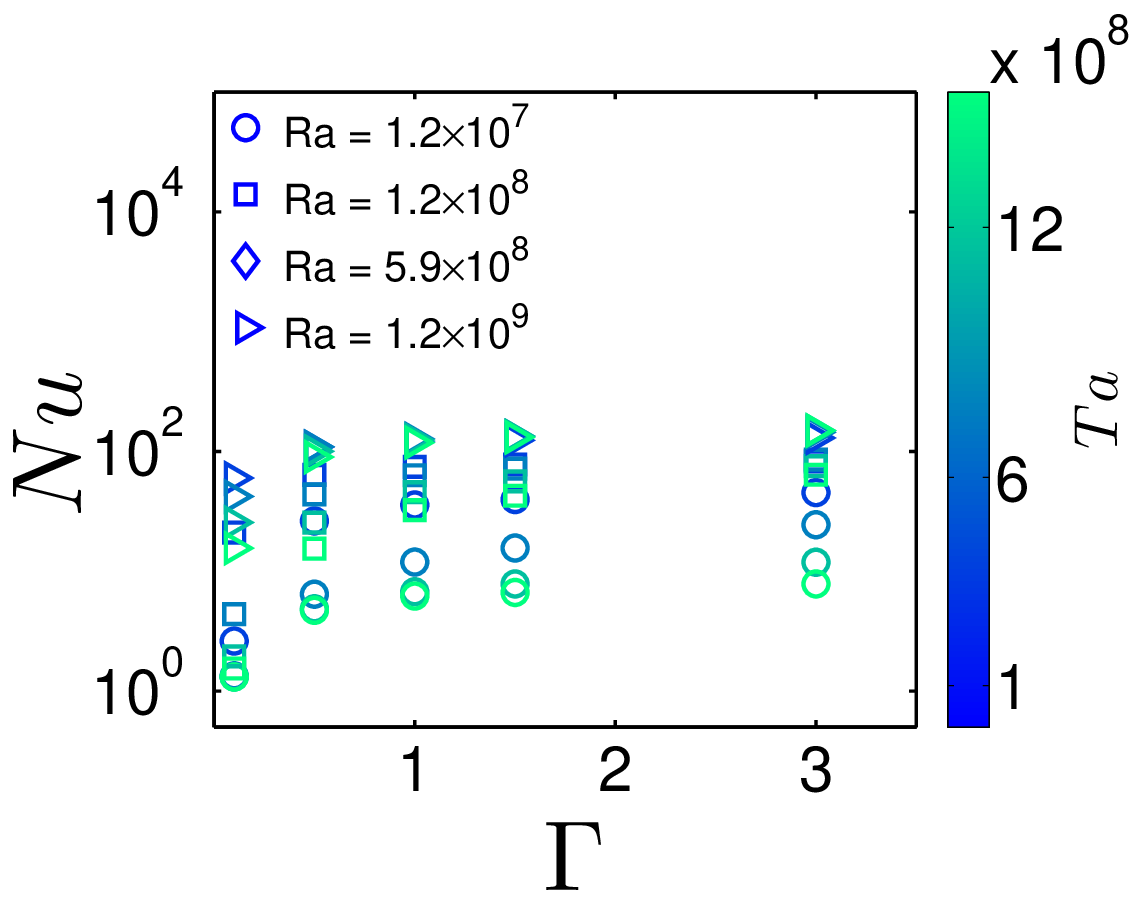}
\label{$t$=82$s$}} &
\subfigure[]{%
\includegraphics[scale=0.32]{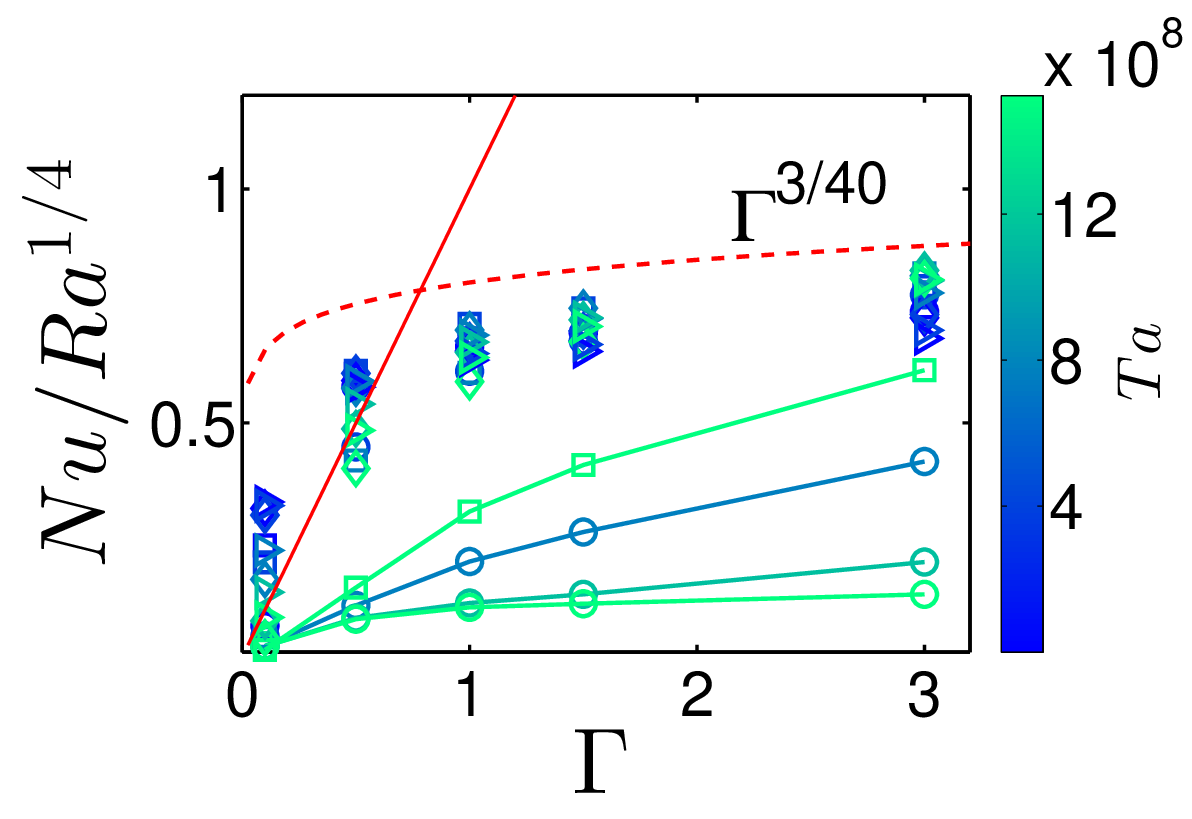}
\label{$t$=102$s$}} 
&\end{tabular}
\caption{\label{Nu_vs_AR_Ta} Nusselt number variation with aspect ratio and $Ra$.  (a) $Nu$ vs $Ra$, (b) $Nu$ vs $\Gamma$, and (c) $Nu/Ra^{1/4}$ vs $\Gamma$}\label{fig:sample1}
\end{figure*}

\begin{figure}
\centering
\subfigure[]{%
\includegraphics[scale=0.35]{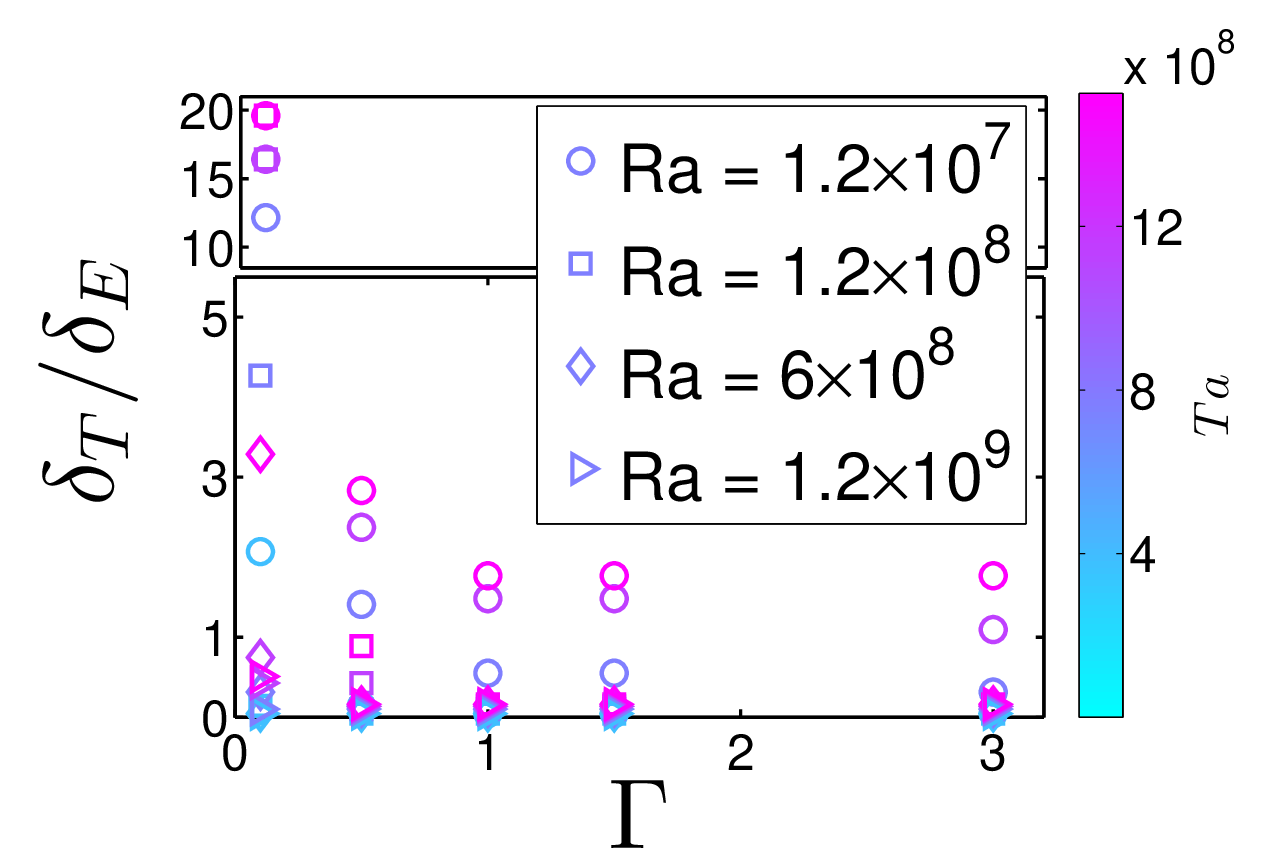}
\label{$t$=102$s$}}
\subfigure[]{%
\includegraphics[scale=0.35]{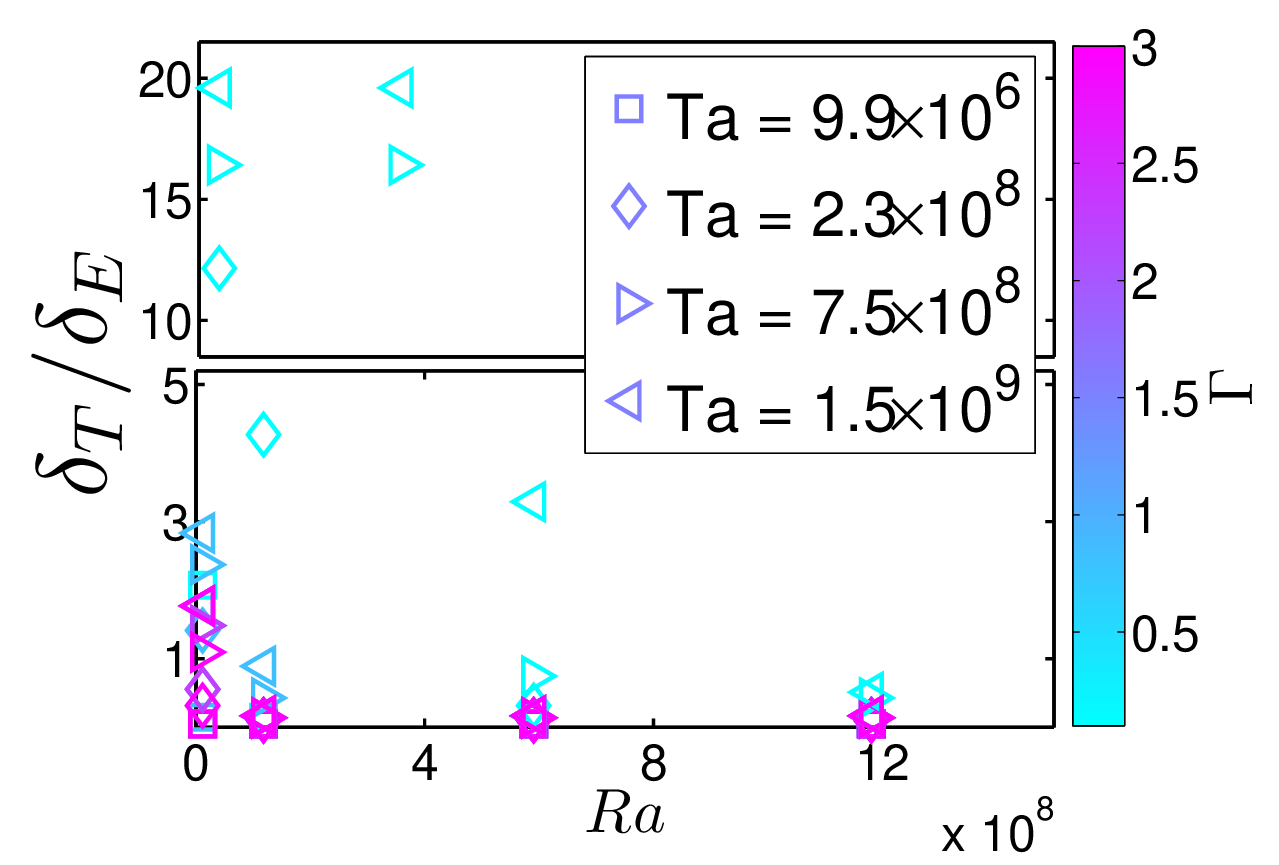}
\label{$t$=82$s$}} 
\caption{\label{BLratio} The ratio of the thermal boundary-layer thickness, \(\delta_T\), to the Ekman boundary-layer thickness, \(\delta_E\), is examined as a function of the governing parameters \(Ra\), \(Ta\), and \(\Gamma\). Figure~(a) shows the variation of \(\delta_T/\delta_E\) with aspect ratio \(\Gamma\), whereas Figure~(b) depicts its dependence on the Rayleigh number \(Ra\). For sufficiently large \(Ra\), the Ekman layer becomes thicker than the thermal boundary layer \((\delta_E > \delta_T)\). Under these conditions, a reduction in the convecting fluid volume, \(Q_f\), has a negligible influence on the overall heat transport, and consequently the Nusselt number remains largely independent of rotation. In contrast, at lower \(Ra\), where \(\delta_E < \delta_T\), increasing rotation progressively constrains the convecting fluid region, thereby reducing convective heat transport efficiency and leading to a decrease in the Nusselt number, \(Nu\).}
\label{fig:sample1}
\end{figure}

Figure~\ref{flow_schematic_simu} illustrates the principal flow features observed in the system. The filled contours represent the normalized vertical-plane velocity,
\[
u_{\text{vert}}^n = \frac{u_{\text{vert}} - u_{\text{vert}}^{\text{min}}}{u_{\text{vert}}^{\text{max}} - u_{\text{vert}}^{\text{min}}},
\] 

where \(u_{\mathrm{vert}}^{\min}\) and \(u_{\mathrm{vert}}^{\max}\) denote the minimum and maximum values of the vertical-plane velocity magnitude,
\[
u_{\mathrm{vert}}=\sqrt{u_r^2+u_z^2},
\]
with \(u_r\) and \(u_z\) representing the radial and axial velocity components in the rotating reference frame. Solid black lines correspond to contours of the normalized temperature field, \(T^{*}\), while blue arrows indicate the direction of fluid motion within the boundary-layer regions.

The flow may be broadly divided into two dynamically distinct regions: a nearly geostrophic bulk interior and thin momentum boundary layers adjacent to the boundaries. Within the boundary layers, residual buoyancy forces generate convective overturning motions in the \(r\)-\(z\) plane. The resulting radial transport is deflected by the Coriolis force, producing an antisymmetric azimuthal circulation characterized by prograde motion near the upper boundary and retrograde motion near the lower boundary. As the rotation rate, quantified by the Taylor number (\(Ta\)), increases, both the velocity magnitude and the thickness of the boundary layers decrease. Consequently, heat transfer between the localized heating zone and the cold inner wall is primarily mediated by convection within the boundary layers, whereas thermal diffusion dominates transport in the weakly circulating geostrophic interior.

\subsection{\textbf{Nusselt number characteristics:}}

To characterize the overall heat-transfer performance of the system, the global Nusselt number, \(Nu\), is employed. This dimensionless quantity represents the ratio of total heat transport to conductive heat transport and is defined as:

\begin{equation}\label{eq:res3_7}Nu=\frac{Q_{total}}{Q_{conduction}}=\frac{Q^{'}}{K \frac{\Delta T}{R}}\end{equation} 
\noindent where, the overall heat flux, denoted as $Q^{'}$, is calculated as,
\begin{equation}\label{eq:Q}Q^{'} = \Big(\int_0^d Q_c^{'} 2\pi r_i dz + \int_{r_0-h}^{r_0} Q_h^{'} 2\pi r dr \Big) \Big/ A_t \end{equation} 
\noindent where $K$ is the thermal conductivity of water, $Q_c^{'}=K \frac{\partial T}{\partial r}\bigg|_{r=r_i}$ is the local heat fluxes, $Q_h^{'}=K \frac{\partial T}{\partial z}\bigg|_{z=0}$ is the local heat fluxes at the hot plate, $A_t=\big(\int_0^d 2\pi r_i dz + \int_{r_0-h}^{r_0} 2\pi r dr \big)$ is the total surface area of the cold wall and the heating plate.

The variation of the Nusselt number, \(Nu\), is presented in Figures~\ref{Nu_vs_Ta_AR} and \ref{Nu_vs_AR_Ta}. The results demonstrate a robust scaling of \(Nu \sim Ra^{1/4}\), although this relationship deteriorates at smaller aspect ratios, \(\Gamma\). In comparison, the influence of rotation on heat transport is relatively weak at moderate and high Rayleigh numbers. However, at sufficiently low \(Ra\), the Nusselt number decreases markedly with increasing Taylor number, \(Ta\), leading to a steeper \(Nu\)--\(Ra\) scaling exponent under stronger rotational effects.

Although rotational forcing disperses the thermal field at large \(Ta\), the momentum field remains largely confined to buoyancy-driven plumes and thin near-wall boundary layers. The results further show that \(Nu\) increases monotonically with the aspect ratio, \(\Gamma\). Larger values of \(\Gamma\) facilitate the formation of larger-scale convective structures, thereby enhancing heat transport between the localized heating zone and the cold wall. The increased horizontal extent allows convective plumes to develop more fully and interact more effectively, promoting hot--cold fluid exchange and consequently increasing the Nusselt number.

For small aspect ratios, the enhancement in heat transfer is pronounced and follows the scaling
\[
\frac{Nu}{Ra^{1/4}} \sim \Gamma,
\]
highlighting the strong influence of geometric confinement on convective transport. In contrast, for \(\Gamma > 1\), the dependence becomes significantly weaker,
\[
\frac{Nu}{Ra^{1/4}} \sim \Gamma^{3/40},
\]
suggesting that the influence of aspect ratio diminishes once the dominant convective structures have fully developed and the bulk flow dynamics become less constrained by geometry. These observations indicate a nonlinear dependence of heat-transfer efficiency on \(\Gamma\), with distinct scaling regimes corresponding to confined and weakly confined convection.

Figure~\ref{BLratio} illustrates the variation of the thermal-to-Ekman boundary-layer thickness ratio, \(\delta_T/\delta_E\), with aspect ratio, \(\Gamma\), for different values of the Taylor number, \(Ta\). For moderate to high Rayleigh numbers, the condition \(\delta_E > \delta_T\) persists even under strong rotational forcing. This behavior arises from the relatively small heating-plate thickness, \(h\), together with the large Prandtl number (\(Pr \gg 1\)), which results in a comparatively thin thermal boundary layer. Under these conditions, the outer portion of the Ekman layer interacts with the bulk fluid, thereby reducing the influence of rotation on the overall heat-transfer process. Consequently, the Nusselt number remains largely insensitive to variations in the convective heat flux, \(Q_f\).

In contrast, when the Rayleigh number is reduced or the rotation rate is increased sufficiently such that \(\delta_E < \delta_T\), a pronounced decrease in \(Nu\) is observed. At high \(Ta\) for a fixed \(Ra\), or equivalently at very low \(Ra\) for a given \(Ta\), the flow velocity weakens substantially. The resulting reduction in buoyancy-driven motion \((\beta g T')\) near the heated and cooled boundaries diminishes the horizontal temperature gradients responsible for driving the circulation. As a consequence, the azimuthal velocity, \(u_\theta\), decreases, leading to a reduction in the convective heat flux, \(Q_f\), and hence a lower Nusselt number.

An additional observation is that the ratio \(\delta_T/\delta_E\) increases as the aspect ratio \(\Gamma\) decreases. This trend enhances the relative influence of rotational constraints on the flow, making the bulk fluid more susceptible to rotational suppression and thereby reducing heat-transfer efficiency. The results therefore demonstrate that the combined effects of \(\delta_T/\delta_E\), \(\Gamma\), and rotation produce a nontrivial dependence of \(Nu\) on the governing geometric and dynamical parameters. While heat transport remains relatively insensitive to rotation at moderate and large aspect ratios, a reduction in \(\Gamma\) increases \(\delta_T/\delta_E\), intensifies rotational suppression, and ultimately leads to an unavoidable decline in the Nusselt number.


\section{\bf{CONCLUSIONS}}

Numerical simulations of two-dimensional axisymmetric flows are conducted using \textsc{ANSYS Fluent} to investigate convection in a bi-directionally forced rotating system. Particular emphasis is placed on understanding the influence of the aspect ratio, \(\Gamma\), on the flow and heat-transfer characteristics over a broad range of Rayleigh (\(Ra\)) and Taylor (\(Ta\)) numbers. The results reveal that the Nusselt number, \(Nu\), exhibits a complex dependence on the governing thermal, rotational, and geometric parameters. For moderate to high \(Ra\), the heat transfer follows an approximate scaling of \(Nu \sim Ra^{1/4}\) and remains largely insensitive to rotation, indicating that buoyancy-driven convection dominates the transport process. In contrast, at sufficiently low \(Ra\) and high \(Ta\), rotational effects suppress convective motion, leading to a marked reduction in \(Nu\) through weakened buoyancy forces and diminished heat flux.

The aspect ratio is found to exert a significant influence on heat-transfer efficiency. At small \(\Gamma\), the Nusselt number increases rapidly with increasing aspect ratio, reflecting the strong impact of geometric confinement on convective transport. As \(\Gamma\) exceeds unity, however, the rate of increase becomes considerably weaker, suggesting that larger domains support fully developed convective structures whose heat-transfer characteristics are less sensitive to further increases in horizontal extent.

The relative thicknesses of the thermal and Ekman boundary layers also play a central role in determining the heat-transfer behavior. When the Ekman boundary layer is thinner than the thermal boundary layer \((\delta_E < \delta_T)\), rotational constraints strongly influence the flow, resulting in a reduction in \(Nu\). Conversely, when \(\delta_E > \delta_T\), interaction between the Ekman layer and the bulk fluid weakens the influence of rotation, rendering the Nusselt number comparatively insensitive to rotational forcing. Overall, the heat-transfer characteristics of the system emerge from a subtle interplay among buoyancy, rotation, and geometric confinement, with the aspect ratio \(\Gamma\) and the boundary-layer thickness ratio \(\delta_T/\delta_E\) serving as key parameters governing the efficiency of convective transport.


\end{document}